\documentstyle[astrobib,referee]{mnv2}

\input psfig

\title[AN ANALYTIC APPROXIMATION TO THE ISOTHERMAL SPHERE]{AN ANALYTIC
APPROXIMATION TO THE ISOTHERMAL SPHERE} 

\author[Priyamvada Natarajan \& Donald Lynden-Bell]
  {Priyamvada Natarajan$^{1}$ \& Donald Lynden-Bell$^{1,2}$ \\
$^{1}$Institute of Astronomy, Madingley Road, Cambridge CB3 0HA \\
$^{2}$Physics Department, The Queens University, Belfast BT7 1NN \\}

\begin{document}
\label{firstpage}
\maketitle

\begin{abstract}
We present a useful analytic approximation to the
solution of the Lane-Emden equation for infinite polytropic 
index - the isothermal sphere. The optimized expression
obtained for the density profile can be accurate to within
0.04\% within 5 core-radii and to 0.1\% within 10 core-radii. 
\end{abstract}
\section{Introduction}

We construct an analytic approximation to the full non-singular
isothermal sphere. The approximations currently in use provide 
a good fit either to the inner regions ($r\,\leq\,2\,{r_{\rm core}}$)
or to the asymptotic behavior. The Lane-Emden equation for the 
gaseous polytrope with polytropic index ${n\rightarrow{\infty}}$ is
identical to that of a self-gravitating 
isothermal sphere. All polytropes with $n\geq$ 5 are infinite and
hence no analytic solutions exist. In a recent paper, \citeN{liu} has
exhaustively examined approximate analytic solutions for polytropes
with general index, where he obtains a solution for the isothermal
case to within $<$ 1\%. In this brief note, we report a
simpler analytic form, accurate to within 0.04\% within 
5 core-radii.

\section{THE ISOTHERMAL SPHERE}

The Lane-Emden equation written in terms of the standard variables
(see \citeNP{Emden}) is,
\begin{eqnarray}
{{d^{2}w} \over {d{\xi^{2}}}}\,+\,{2 \over {\xi}}
{dw \over d\xi}\,=\,{e^{-w}},
\end{eqnarray}
where $w\,=\,\ln\,{\rho_0}/{\rho}$; ${\xi\,=\,{r/r_{0}}}$; $r_{0}^2
= \sigma^2/{4 \pi G {\rho_0}}$; $\sigma$ is the constant velocity
dispersion and $r_{0}$ the core radius. 

The solution for the density distribution can be expanded into a series,
\begin{eqnarray} 
{{\rho} \over {\rho_{0}}}\,=\,1\,-\,({1 \over 6}){\xi^{2}}\,+\,({1 \over 45}){\xi^{4}}\,-\,...
\end{eqnarray}
The approximation that is used often in the literature (which we
denote by the subscript (a) henceforth) follows from truncating
equation (2) at the second order. 
\begin{eqnarray}
{\rho(\xi)_{a}}\,=\,{{1 \over {1+{{\xi^2} \over
6}}}}.
\end{eqnarray}  
While the mathematical form is simple, this profile overestimates 
mass outside ${r_{0}}$ and differs from the exact solution by a 
factor of 3 as ${r\rightarrow{\infty}}$.

As an ansatz, we attempt the following analytic form for the
approximation:
\begin{eqnarray}
{\rho(\xi)_{\rm approx}}\,=\,{{A \over {a^2+{\xi^2}}}}\,-\,{{B \over {b^2+{\xi^2}}}},
\end{eqnarray}
where A, B, $a^2$ and $b^2$ are to be determined. 
The solution obtained (see Appendix for details) is given by,
\begin{eqnarray}
{\rho(\xi)_{b}}\,=\,{{5 \over {1+{{\xi^2} \over
10}}}}\,-\,{{4 \over {1+{{\xi^2} \over 12}}}}.
\end{eqnarray}
This expression is correct asymptotically and is accurate to within ${1\%}$
up to ${\xi\,=\,5}$. Preserving the general form, this approximation 
can be optimized further yielding,
\begin{eqnarray}
{a^2}\,=\,{[10-({{5 \over 11}\delta-{2 \over 11}\epsilon})]}\,\,\,;\,\,\,{b^2}\,=\,{[12-({{9 \over 11}\delta+{3 \over 11}\epsilon})]}\,; \nonumber
\end{eqnarray}
\begin{eqnarray}
{A \over {a^2}}\,=\,{(5+\epsilon)}\,\,\,;\,\,\,{B \over
{b^2}}\,=\,{(4+\epsilon)},
\end{eqnarray}
where the additional parameters $\epsilon$ and $\delta$ have now 
been chosen (see Appendix for details) to optimize the degree of 
accuracy in terms of agreement with the full exact solution.  
This solution is plotted in Figures 1 and 2. The mathematical
form of the optimized solution is convenient appropos Abel inversion
since it can be inverted analytically. 
\begin{figure}
\psfig{figure=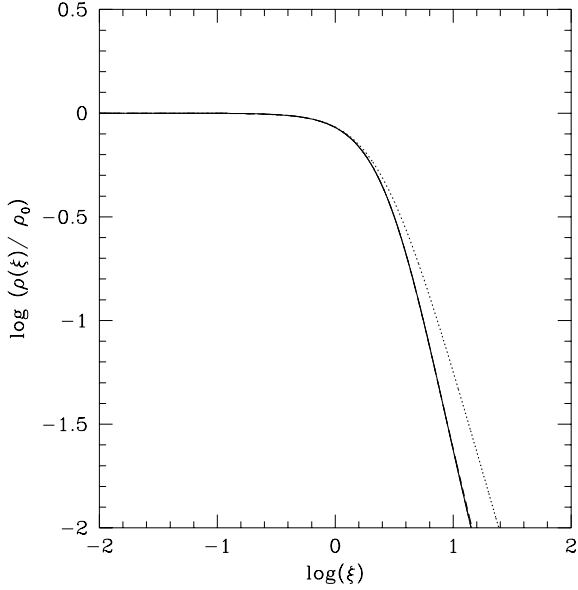,width=0.5\textwidth}
\caption{VARIOUS APPROXIMATIONS TO THE ISOTHERMAL SOLUTION: solid
curve - exact solution, dotted curve - approx(a), dashed curve - approx(b), 
long-dashed curve - approx(c)}
\end{figure}
The corresponding projected quantities for our formula -  the surface
density and mass are easily computed and have the following
analytic forms: 
\begin{eqnarray}
\Sigma(R)\,=\,{A\pi \over {\sqrt{a^2+R^2}}}\,-\,{B\pi \over
{\sqrt{b^2+R^2}}},
\end{eqnarray}
where R is the projected radius. These spherical models can be easily 
generalized to describe elliptical mass distributions as well, akin to
those proposed by \citeN{kovner93}.
\begin{eqnarray}
{M_{\rm 3D}}(\xi)\,=\,4\pi\,[\,{A\,a\,({{\xi}
\over a}\,-\,{\tan^{-1}({{\xi} \over a})})}\,-\,{B\,b\,({{\xi} \over b}\,-\{\tan^{-1}({{\xi} \over b}))}],
\end{eqnarray}
The potential on the plane corresponding to the surface density
$\Sigma$ is,
\begin{eqnarray}
\phi_{2D}\,=\,A\pi[{\sqrt{a^2+R^2}}-{a \ln R}-{a \ln
(a^2+a{\sqrt{a^2+R^2}})}]\\ \nonumber\, -\,
B\pi[{\sqrt{b^2+R^2}}-{b \ln R}-{b \ln(b^2+b{\sqrt{b^2+R^2}})}].
\end{eqnarray}
\begin{figure}
\psfig{figure=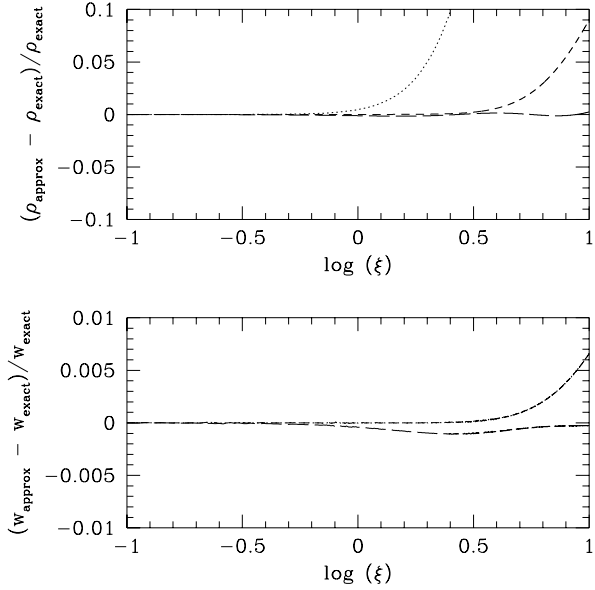,width=0.5\textwidth}
\caption{DIFFERENTIAL ERRORS: (i) TOP PANEL - density (dotted
curve - approx(a), dashed curve - approx(b), long-dashed curve - approx(c))
and (ii) LOWER PANEL - the potential (dotted
curve - approx (a), dashed curve - approx(b), long-dashed curve -
approx(c)) (the dotted curve and dashed curve are coincident)}
\end{figure}
\begin{figure}
\psfig{figure=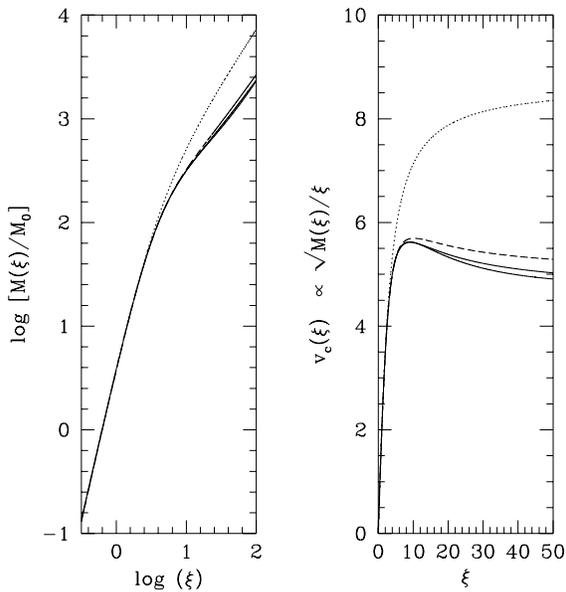,width=0.5\textwidth}
\caption{(i) LEFT PANEL: MASS ENCLOSED WITHIN $r$ (solid curve - exact solution,
dotted curve - approx(a), dashed curve - approx(b), long-dashed curve
- approx(c))
and (ii) RIGHT PANEL: circular velocity profiles (solid curve - exact solution,
dotted curve - approx(a), dashed curve - approx(b), long-dashed curve
- approx(c)) }
\end{figure}
These projected quantities are of interest in many physical
problems - for instance, in the context of modelling gravitational 
lensing observations. The two primary effects producing by lensing are
the isotropic manginification and the anisotropic shear. The
magnification $\kappa$ produced by the potential is given by,
\begin{eqnarray}
\kappa(R)\,=\,{\kappa_0}[\,{{A\pi} \over {\sqrt{a^2+R^2}}}\,-\,{{B\pi} \over 
{\sqrt{b^2+R^2}}}\,]
\end{eqnarray}
where ${\kappa_0}\,=\,{1 \over {\Sigma_{\rm crit}}}$; and $
{\Sigma_{\rm crit}}$ is the critical surface density given the geometrical
configuration of the source, lens and the observer. And the induced
shear $\gamma$ is,
\begin{eqnarray}
\gamma(R)\,&=&\,\nonumber 
\kappa_0[\,-{{A\pi} \over \sqrt{R^2 + a^2}}\,
        +\,{{2A\pi} \over R^2}(\sqrt{R^2 + a^2}-a)\,\\
\nonumber 
&+&\,{{B\pi} \over {\sqrt{R^2 + b^2}}}\,-\,
{{2B\pi} \over R^2}(\sqrt{R^2 + b^2}-b)\,].\\
\end{eqnarray}

\section{Conclusions}

The analytic approximation presented above is potentially useful in
the context of many physical problems and is particularly useful since
the projected quantities have simple analytic forms. We also point out
that within 5 core-radii, while the analytic approximation to the 
isothermal sphere currently in use systematically over-estimates the 
mass enclosed our formula is accurate to within 0.04\%.

\section{APPENDIX}

\subsection{Fitting to the exact solution}

Starting with our ansatz for the functional form,
\begin{eqnarray}
{\rho(\xi)_{\rm approx}}\,=\,{{A \over {a^2+{\xi^2}}}}\,-\,{{B \over {b^2+{\xi^2}}}}.
\end{eqnarray}
Expanding the above as,
\begin{eqnarray}
{\rho(\xi)_{\rm approx}}\,=\,{{A \over {a^2}}({1 \over {1+{{\xi^2} \over {a^2}}}})}\,-\,{{B \over {b^2}}({1 \over {1+{{\xi^2} \over {b^2}}}})},
\end{eqnarray}
\begin{eqnarray}
{\rho(\xi)_{\rm approx}}\,=\,{{A \over {a^2}}(1\,-\,{{\xi^2} \over
{a^2}}\,+\,{{\xi^4} \over {a^4}}\,+\,...)}\, \\ \nonumber - \,\,\,\,\,\,
{{B \over {b^2}}(1\,-\,{{\xi^2} \over {b^2}}\,+\,{{\xi^4} \over {b^4}}\,+\,...)},
\end{eqnarray}
Comparing terms to corresponding orders in ${\xi}$ in equations 1 and 2, we
obtain the following system of equations:
\begin{eqnarray}
{A \over {a^2}}\,-\,{B \over {b^2}}\,=\,1\,\,;\,\,\,{A \over {a^4}}\,-
\,{B \over {b^4}}\,=\,{1 \over 6}\,\,;\,\,\,{A \over {a^6}}\,-
\,{B \over {b^6}}\,=\,{1 \over 45}.
\end{eqnarray}
Requiring the asymptotic behavior as $\xi\,\to\,\infty$ to match up
to the full solution, we obtain the additional equation to close the
above system of 4 simultaneous equations in 4 unknowns.
\begin{eqnarray}
A\,-\,B\,=\,2,
\end{eqnarray}
we obtain an exact solution!,
\begin{eqnarray}
{\rho(\xi)_{b}}\,=\,{{5 \over {1+{{\xi^2} \over
10}}}}\,-\,{{4 \over {1+{{\xi^2} \over 12}}}}.
\end{eqnarray}

\subsection{Optimized approximation}

We introduce additional parameters $\epsilon$, $\delta$, x and y
and constrain their values to the desired degree of accuracy
as follows,
\begin{eqnarray}
{\rho(\xi)_{c}}\,=\,{{{5\,+\,\epsilon} \over {1+{{\xi^2} \over
{10\,-\,{x\epsilon}}}}}}\,-\,{{{4\,+\,\epsilon} \over {1+{{\xi^2}
\over {12\,-\,{y\epsilon}}}}}}.
\end{eqnarray}
We require agreement asymptotically as ${\xi \rightarrow \infty}$,
therefore simplifying the equation above for large $\epsilon$,
\begin{eqnarray}
(5+\epsilon)(10-x\epsilon)\,-\,(4+\epsilon)(12-y\epsilon)\,=\,2,
\end{eqnarray}
ignoring terms of $O({\epsilon^{2}})$, 
\begin{eqnarray}
x\,=\,({{5 \over 9}y}\,-\,{1 \over 3}).
\end{eqnarray}
Substituting the above back into equation (16),
\begin{eqnarray}
{\rho(\xi)_{c}}\,=\,{{5\,+\,\epsilon} \over {1+{{\xi^2} \over
{10\,-\,{x\epsilon}}}}}\,-\,{{4\,+\,\epsilon} \over {1+{{\xi^2}
\over {12\,-\,({{1 \over 2}+{{5 \over 4}x\epsilon}})}}}}.
\end{eqnarray}
Now expanding the above two terms on the RHS into a series
and matching the corresponding terms of the same order in $\xi$
to $O({\epsilon})$ and matching asymptotically to $\delta$ degree 
of accuracy we have,
\begin{eqnarray}
y\,=\,({3 \over 11}\,+\,{9 \over 11}{\delta \over \epsilon}).
\end{eqnarray}
For an RMS error of 0.04\% to ${\xi\,=\,5}$, we find
${\epsilon\,=\,-0.07}$ and ${\delta\,=\,-0.269}$. If we do
not treat $\epsilon$ as small and do not demand the correct
asymptotic solution at large $\xi$, we can obtain a solution
with RMS error of 0.1\% at ${\xi\,=\,10}$. This is
${\epsilon\,=\,-0.635}$ and ${\delta\,=\,-0.6}$ which is used 
in the plots. In this  note, we have approximated the full
untruncated isothermal sphere and have not addressed the problems
of truncating it.

\section*{ACKNOWLEDGMENTS}

PN acknowledges funding from the Isaac Newton Studentship 
and Trinity College at the University of Cambridge. 

\bibliography{mnrasmnemonic,refs}
\bibliographystyle{mnrasv2}
\end{document}